\documentclass[%
reprint,
showpacs,preprintnumbers,
amsmath,amssymb,
aps,
pra,
]{revtex4-1}
\usepackage{float}
\usepackage{graphicx}
\usepackage{dcolumn}
\usepackage{bm}
\usepackage{epstopdf}
\usepackage{float}
\usepackage{xcolor}
\usepackage{hyperref}

\begin{document}
	
	
	\title{Dissociative electron attachment to pulsed supersonic O$_2$ jet : Violation of $\Sigma^{+} \rightleftharpoons \Sigma^{-}$ selection
		rule and dependence on carrier gas proportion}
	%
	\author{Irina Jana}
	\author{Varun Ramaprasad}
	\author{Dhananjay Nandi}%
	\email{dhananjay@iiserkol.ac.in}
	\affiliation{Department of Physical Sciences,\\\\
		Indian Institute of Science Education and Research Kolkata, Mohanpur 741246, India\\
	}

	\begin{abstract}
		
		The formation of $O^{-}$ and $O_{2}^{-}$ ions via dissociative
		electron attachment to a pulsed supersonic jet of $O_{2}$ molecules
		containing weakly bound small van der Waals clusters seeded in a beam
		of argon is reported. The energy dependence of the $O^{-}$ and
		$O_{2}^{-}$ yield exhibits three peaks near 7, 11 and 16 eV incident
		electron energies. The 7 eV peak arises from the $^{2}\Pi_{u}$ state
		of $O_{2}^{-}$ whereas, the 11 and 16 eV peaks are ascribed to two
		distinct resonance states: $ ^{2}\Sigma_{g}^{+} $ and
		$^{2}\Sigma_{u}^{+}$ states of $O_{2}^{-}$, respectively, via a
		violation of the $\Sigma^{+} \rightleftharpoons \Sigma^{-}$ selection
		rule. The dependence of the cross-section of these two new peaks
		at $\sim$11 and $\sim$16 eV on the proportion of the carrier gas is also
		investigated and an optimum proportion has been observed
		experimentally which gives the lowest temperature of 14.86 K and
		highest Mach number of 72.31 for the pulsed supersonic jet. 
		
	\end{abstract}
	
	
	\maketitle
	

\section{Introduction}
Clusters of small molecules and atoms are important intermediates
between the gas phase and the condensed phase. These are of importance
in many chemical and physical processes starting from living tissues
to atmospheric chemistry and hence is the topic of interest since the past few decades \cite{helium-clusters,prl-supersonic-laser,prl-water-cluster}. We study here the electron attachment process with isolated oxygen molecules and small van der Waals
clusters of oxygen formed in pulsed supersonic jet of oxygen
molecules. Electron capture by an effusive oxygen beam is known to
produce a resonance at 6.5 eV incident electron energy following the
reaction \cite{jaffke1992,henderson1969,mark1985,tashiro2006}:
\begin{equation}
	 O_{2}\ (^{3}\Sigma_{g}^{-})\ +\ e^{-}\ \rightarrow\ (O_{2}^{-})^{*}\
	(^{2}\Pi_{u})\ \rightarrow \ O^{-}\ (^{2}P)\ +\ O\ (^{3}P)
\end{equation}
 While electron attachment with oxygen clusters produces mainly two
homologous series:$(O_{2})_{n}^{-}$ and $(O_{2})_{n}O^{-}$ and can be
represented as :
\begin{equation}
	(O_{2})_{x} \ +\ e^{-} \ \rightarrow\ (O_{2})_{x-1}
	O^{-}\ +\ O
\end{equation}
which may be followed by subsequent isomerization \cite{mark1985}.\\

Marc \textit{et al.} studied the electron attachment to oxygen clusters for
the two homologous series $(O_{2})_{n}^{-}$ and $ (O_{2})_{n}O^{-}$
by expanding the gas through a 10 $\mu m$ nozzle \cite{mark1985}. Other
than the 6.5 eV peak for $O^{-}$ ion, they also reported the formation
of a peak near 0 eV, forming due to a non-dissociative attachment to
$O_{2}$, while the dimer $ O_{2}^{-}$ ion is seen to produce a peak $\sim$7 eV along with the high cross-section 0 eV peak. Matejcik \textit{et al.} also
reported vibrationally resolved electron attachment to $(O_{2})_{n}$
clusters from 0 to 2 eV incident electron energies expanding the $O_{2}$ gas using a 20 $\mu m$ nozzle \cite{matejcik1996}. Back in 1987,
Azria et al. reported the formation of $O^{-}$ ions from dissociative
electron attachment (DEA) in electron stimulated desorption (ESD) to
three-layer-thick thin film of $O_{2}$ at two resonances near 7 and
13 eV incident electron energies. The 13 eV peak was ascribed due to
two distinct $O_{2}^{-}$ resonance states : $ ^{2}\Pi_{u}$ and
$^{2}\Sigma_{g}^{+} $, via a violation of the selection rule $\Sigma^{-}
\rightleftharpoons \Sigma^{+}$ \cite{Azria-selectionrule}. Illenberger in his review of electron attachment process to different molecular clusters claimed that DEA to $(O_{2})_{n}$ clusters produced two new
distinct peaks near 8.3 and 14.5 eV incident electron energies for the
dimer $(O_{2})^{-}$ ion and also $O^{-}$ ion \cite{illenberger1992}.\\

We report in this article a clear observation of two new distinct resonance peaks at $\sim$11 and $\sim$16 eV incident electron energies along
with the 7 eV peak, from $O^{-}$ and $O_{2}^{-}$ ions formed by
dissociative electron attachment to a pulsed supersonic jet of $O_{2}$
molecules  containing $(O_{2})_{n}$ clusters seeded in a beam of argon.\\


We also investigated for the first time the dependence of the cross -
section of these new symmetry forbidden resonance peaks, at $\sim$11 and $\sim$16 eV incident electron energies, on the proportion of the seeder
gas. The change in the pulsed supersonic beam temperature and Mach
number with change in argon proportion is also investigated for the
first time and the results are explained.

\section{Experimental setup}
\begin{figure}[h]
	\centering	
	\includegraphics[width=1\linewidth]{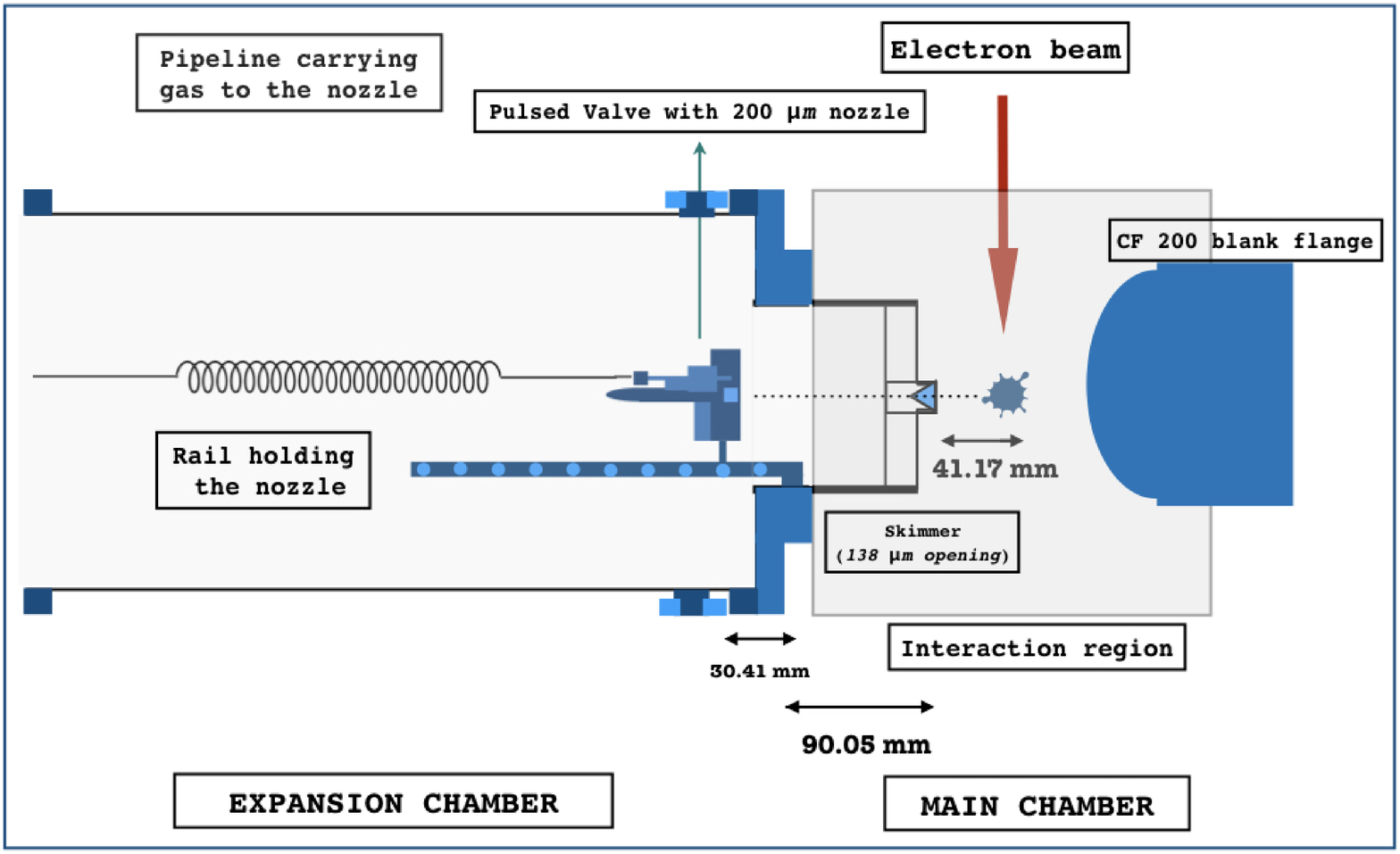}
	\caption{\label{setup} Schematic with appropriate dimensions showing the pulsed valve holding the nozzle on the rail. }
\end{figure}
In order to have a supersonic jet of molecules, a high pressure gas is passed through a nozzle into a low pressure environment where it expands quasi-statically and adiabatically. In this set-up, the Amsterdam Cantilever Piezo Valve (ACPV) is placed in an expansion chamber with a 200 $\mu m$ nozzle which can be controlled by an electronic driver unity (EDU). The ACPV can be operated both in continuous and pulsed (0 - 5 kHz) mode giving either a continuous or pulsed  supersonic jet of molecules \cite{irimia2009}.

A pulsed supersonic jet reduces the cost of pumping by eliminating the need for a catcher chamber which is necessary for a continuous supersonic jet. In the current setup, a pulsed supersonic jet of 37 $\mu s$ (FWHM) is produced by triggering the EDU with an input TTL pulse of 1.5 kHz using a master pulse generator. Another pulse output of 200 ns pulse width from the same master pulse generator goes to the input channel of an Ortec G8020 delay generator. All other necessary pulses are taken from different outputs of this delay generator to have a constant reference delay with respect to the molecular beam pulse. A 138 $\mu m$ skimmer, separating the two chambers, is placed inside the isentropic region, termed as the \textit{zone of silence}, to extract the centreline beam consisting of the coldest jet of molecules. The expansion chamber containing the pulsed valve is pumped with the help of a TC400 turbo molecular pump while, the main chamber containing the electron gun, Faraday cup and the time-of-flight (TOF) mass spectrometer is pumped with the help of a TC700 turbo molecular pump. Both the chambers are maintained at a base pressure of $\sim10^{-8}$ mbar with no gas and a pressure of $\sim10^{-4}$ mbar in the expansion chamber and $\sim10^{-7}$ mbar in the main chamber with the gas inflow. To ensure the skimmer lies within the \textit{zone of silence} and away from the Mach disc, the ACPV is placed on a rail allowing the distance between the nozzle and the skimmer to be changed. Details of the structure of a cold supersonic jet has already been explained in details earlier in many works \cite{Cameron,zhang2006}. 
\begin{figure}[h]
	\centering	
	\includegraphics[width=1\linewidth]{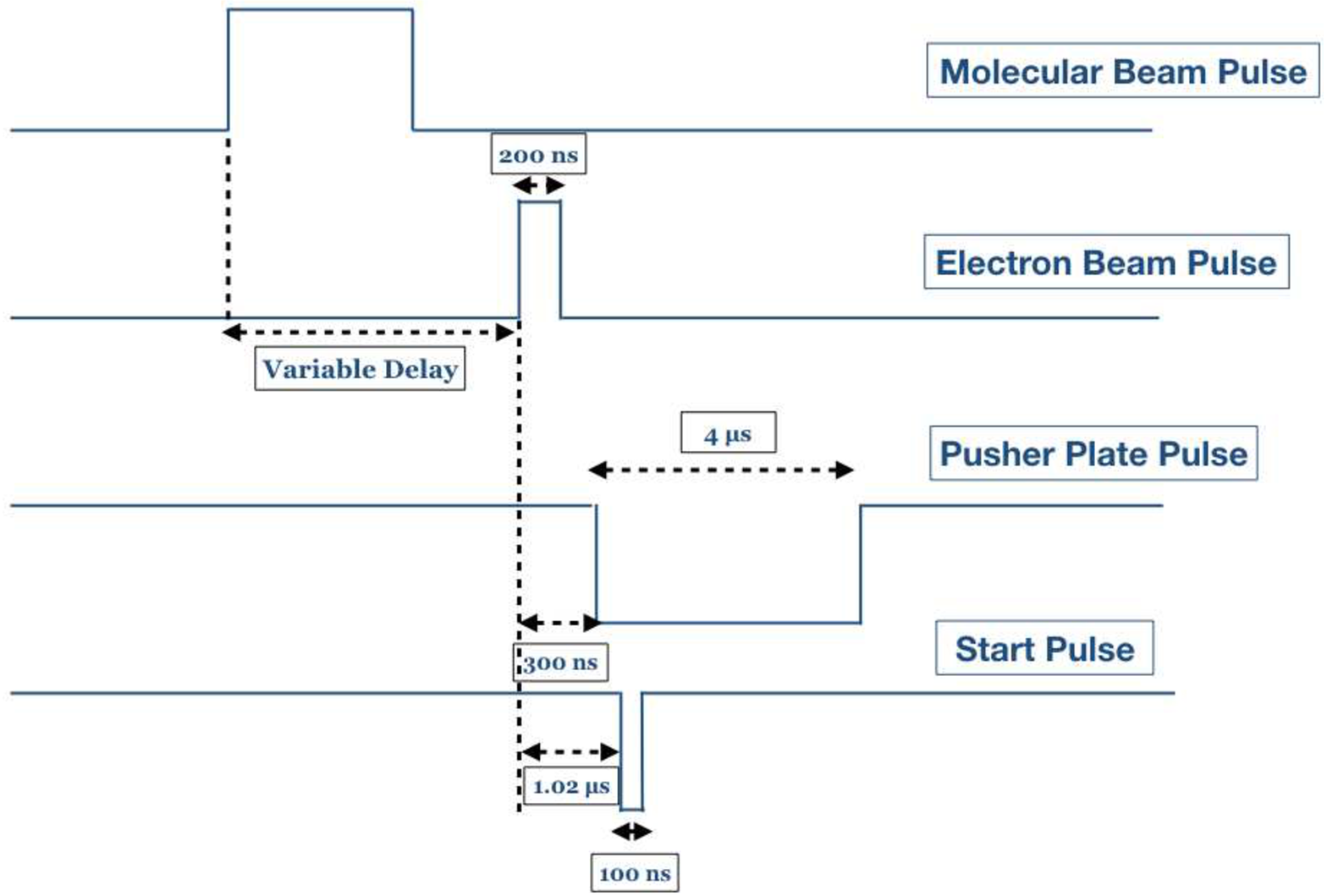}
	\caption{\label{pulse-diagram} Pulse diagram showing different pulses (not to scale).}
\end{figure}
Once the pulsed centreline beam enters the interaction region (in the main chamber), it is crossed with a pulsed electron beam emitted from a home-made electron gun. \\

The electron gun has a tungsten filament heated with a 2.1 A constant current source which emits electrons through thermionic emission. The electron gun is pulsed with a 1.5 kHz TTL pulse of 200 ns pulse-width coming from the output of channel 1 of the G8020 delay generator. The pulsed supersonic jet takes a finite time to travel from the point of origin i.e, the nozzle to the interaction region. The delay between electron gun pulse and molecular beam pulse is varied from the delay generator to ensure proper synchronization between the two pulses. The electron gun current is constantly monitored with the help of a Faraday cup placed below the electron gun kept at a positive voltage of 40 V.\\

The crossed pulsed electron beam and supersonic jet collide under single collision condition producing negative ions in all 4$\pi$ directions, thus forming the \textit{Newton sphere} of negative ions. The \textit{Newton spheres} formed are pushed from behind by the pusher plate having a pulsed negative voltage of -20 V. The pusher plate is pulsed with a 4 $\mu s$ pulse having a 300 ns delay with respect to the electron gun. This delay ensures better extraction of the negative ions. The Newton spheres of negative ions are then collimated with the help of the grounded puller plater and the lens electrode having a positive polarity of 18 V. The Newton spheres then finally enter the 110 mm long field-free flight tube having a positive polarity of 100 V, where they expand freely and finally reach the MCP based detector. The detector consists of three micro-channel plates (MCP’s) placed in a Z-Stack configuration. The TOF of each ion reaching the detector and the number of counts at a defined incident electron energy are recorded with a TAC unit using LabVIEW. Details of the main chamber containing the electron gun, Faraday cup and time-of-flight mass spectrometer have already been discussed before \cite{nag2015}.

\section{Formation of an optimum jet}
Ideal gas condition of stationary flow is assumed throughout where a high pressure gas is allowed to pass through a small aperture and expand into a low pressure region where it undergoes an isentropic wall-free expansion without condensation.\\ 

The most important jet equation relating the stagnation enthalpy $H_{0}$, kinetic energy of the directed mass flow (E) and rest enthalpy (H) can be written as: 
\begin{equation}
H_{0} \ =\  E\  +\  H
\end{equation}
where
\begin{equation}
E \ =\  \dfrac{1}{2}\ m\ u^{2}
\end{equation}
and 
\begin{equation}
H\ = \ U\ +\ PV
\end{equation}
where \textit{u} is velocity of the jet, \textit{U} is internal energy, \textit{P} is pressure and \textit{V} is volume \cite{zhang2006,haberland1985}.\\

The temperature \textit{$T_{0}$} and \textit{T} before and after the aperture can be related using the specific heat at constant pressure (\textit{$c_{P}$}) using ideal gas behavior as:
\begin{equation}
	c_{P}\ T_{0}\ =\ c_{P}\ T\ +\ \dfrac{1}{2}\ m\ u^{2}
\end{equation}
or 
\begin{equation}
 \dfrac{T}{T_{0}}\ =\ \dfrac{1}{[1\ +\ \frac{\gamma\ -\ 1}{2}\ M^{2}]}
 \end{equation}
Again, for an ideal, isentropic gas expansion with constant $\gamma$, we can write:
\begin{eqnarray}
	\frac{P}{P_{0}}\ =\ (\frac{T}{T_{0}})^{\frac{\gamma}{\gamma\ -\ 1}}\ =\ [1\ +\ \frac{\gamma\ -\ 1}{2}\ M^{2}]^{\frac{-\gamma}{\gamma\ -\ 1}} \nonumber\\
	\frac{\rho}{\rho_{0}}\ =\ \frac{n}{n_{0}}\ =\ (\frac{T}{T_{0}})^{\frac{1} {(\gamma\ -\ 1)}}\ =\ [1\ +\ \frac{\gamma\ -\ 1}{2}\ M^{2}]^{\frac{-1}{\gamma\ -\ 1}} \nonumber
\end{eqnarray}
where \textit{$P_{0}$}, \textit{P} are the pressures before and after the aperture, \textit{$\rho,\ \rho_{0}$} are densities before and after the aperture and \textit{$n, n_{0}$} are number densities before and after the aperture.\\

The above equations give one a clear quantitative idea about the Mach number (\textit{M}) and local temperature (\textit{T}) along the streamline of the expanding gas jet. Consequently, all thermodynamic variables can be computed in the supersonic jet.\\

The end of the isentropic region, known as the ‘\textit{zone of silence}’ is  marked by the Mach disc \cite{zhang2006}. The distance of the Mach disc from the nozzle is related to the nozzle diameter as: 

\begin{equation}
	x_{M}\ =\ d\ (0.67)\ \sqrt{\frac{P_{0}}{P}}
\end{equation}

where \textit{$x_{M}$} is the Mach disc location and \textit{d} is the opening of nozzle. The location of the Mach disc was calculated to be 857.47 mm with a 200 $\mu$m nozzle, \textit{$ P_{0}$} was $\sim$ 4 bar and \textit{P} was $\sim$ $10^{-4}$ mbar. It was ensured that the skimmer, to extract the centreline beam, was placed inside the \textit{zone of silence} stretching to a distance of 847.47 mm from the nozzle. \\

In the present chamber, the pulsed valve fitted on a rail can be moved to different positions changing the nozzle-to-skimmer distance \ref{setup}. In order to investigate the effect of nozzle-to-skimmer distance variation on the observed beam interaction, the pulsed valve holding the nozzle was manually placed at 3 different positions and the beam intensity of $O^{-}$ and $O_{2}^{-}$ ions were studied keeping the stagnation pressure fixed at 5 bar and expansion chamber pressure at $7 \times 10^{-4}$ mbar for all 3 positions. As the nozzle-to-skimmer distance was changed from screw points 1 to 3, it could be observed that the $O_{2}^{-}$ cluster ions counts reached an optimum at screw point 2. As the pulsed valve was moved to the first screw point on the rail, which is the closest to the skimmer, high jet densities formed at the nozzle reach the skimmer. As a result, the $ O^{-}$ ions counts increased strikingly while the formation of $O^{-}_{2}$ ion decreased to a great extent. The strong barrel shock waves produced at the boundaries of the high density jets might not get attached to the skimmer and thus combine with the Mach disc which lies perpendicular to the direction of supersonic flow scattering the jet particles. This results in a highly degraded supersonic beam which hinders the formation of the weakly bound van der Waals’ clusters  i.e, the $O_{2}^{-} $ ions \cite{im2009-PRL-shockwavs}.  \\

On the other hand, when the pulsed valve was placed on the third screw position, the number of counts of $O_{2}^{-}$ ions were found to increase considerably in comparison with the $O^{-}$ ions but the total number of counts of both $O_{2}^{-}$ and $O^{-}$ ions was observed to be very low resulting in the remarkably low beam intensity at the interaction region. In fact, when the distance between the nozzle and skimmer is large, the intensity of gas particles varies following the classical scattering equation:

\begin{equation}
	 I \ =\ I_{0} \ e^{m\ \sigma_{eff} \ z}
\end{equation} 
 
where, \textit{$I_{0}$ }and \textit{I }are the ideal and observed beam intensities, respectively; \textit{m} is the background number density in the expansion chamber; \textit{z} is the length over which scattering occurs and $\sigma_{eff}$ is the effective scattering cross section \cite{levine2005,Cameron}. \\

Whereas, when the pulsed valve was placed at the second screw point, two noticeably different peaks in the mass spectra were observed at masses 16 and 32 a.m.u. corresponding to $O^{-}$ and $O_{2}^{-}$ ions with sufficiently high beam intensities.  This supports the well known fact that for an  optimum nozzle-to-skimmer distance, the beam intensity reaches the maximum. Shifting the nozzle, closer or further away from the optimum distance, results in a remarkable decrease in the beam intensity \cite{Cameron}. \\

In order to have proper synchronization between he pulsed molecular beam and pulsed electron beam, it is necessary to know the time taken by the supersonic jet from the nozzle to reach to the interaction region. It has to be assured that when the jet has reached the interaction region then only the electron beam has been put on. The pusher plate pulse and the start pulse are well synchronized with the electron gun pulse. Schematic of all the necessary pulses used in the experiment are shown in Figure \ref{pulse-diagram}. To find out the exact time taken by the jet to cover the distance between its point of origin and the interaction region, the delay between the molecular beam pulse and the electron gun pulse was varied slowly using the DGB 35 digital delay pulse generator from Stanford Research Systems and the $O^{-}$ ion counts were noted at 6.5 eV incident electron energy using LabVIEW, since $O^{-}$ is well known to have a resonant peak at 6.5 eV. The delay for which a maximum count was observed was noted.    


\section{Cluster formation with a seeded beam:}

\begin{figure}[h]
	\centering	
	\includegraphics[width=1.1\linewidth]{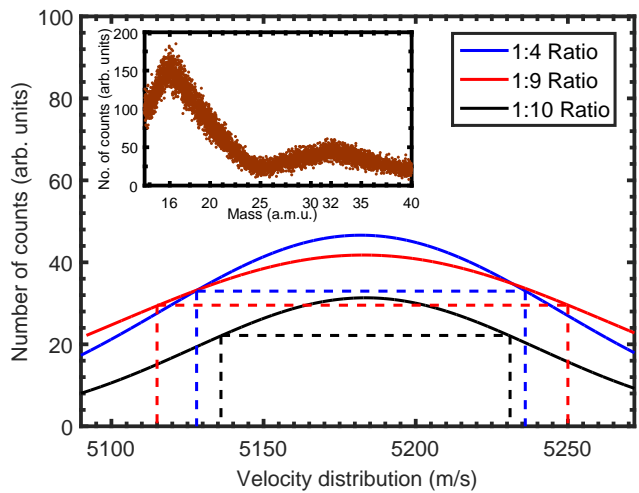}
	\caption{\label{vel-compared} Velocity distribution fitted curves using Equation \ref{vel-dist} for different O$_2$:Ar ratios 1:4 (Blue line), 1:9 (Red line) and 1:10 (Black line) compared together.}
\end{figure}

For a mono-atomic gas final velocity of the supersonic jet can be
expressed as \cite{illenberger1992}:

\begin{equation}
	v_{t}\ =\ \sqrt{\frac{5\ k\ T_{0}}{M_g}}
	\label{mass}
\end{equation}

Here \textit{$M_g$} is the mass of the gas. This terminal velocity ($v_t$) is denoted by the peak position of the narrow velocity
distribution which is a characteristic feature of the supersonic jet.\\
%

The situation becomes a little complicated for polyatomic
molecules as $\gamma$ $\sim$ 1, making the cooling process less effective. To address this problem, one can use an inert carrier gas like He or Ar. The molecules are diluted in small
proportion in a stream of inert carrier gas, argon in this case, such
that the mass (\textit{$M_g$}) in Equation \ref{mass} is now the mass of the carrier gas Ar.
The carrier gas acts as a refrigerant taking away the excess heat from
the jet and thus helping it to polymerize. Thus the use of a carrier
gas makes the cooling of the jet more effective which in turn, results
in better cluster formation.\\

After the optimum distance between the nozzle and the skimmer is
identified, a beam
of oxygen molecules diluted with Ar is then
allowed to expand through the nozzle in the expansion chamber. The
ratio of oxygen and argon has been varied over a wide range and the
mass spectra and ion-yield curves for both $O^{-}$ and $ O_{2}^{-}$
ions have been noted. The difference in the above mentioned curves
with change in argon proportion are then investigated and reported
here.\\

The mass spectra with $O_{2}:Ar$=1:10 is shown in the inset of
Figure \ref{vel-compared}. With the change of Ar proportion the mass spectra do not show much visible change while, the number of counts and FWHM change. The first peak of the mass spectra corresponds to $O^{-}$ ions (16 a.m.u) while the second
peak corresponds to $O_{2}^{-} $ions (32 a.m.u). \\

The TOF spectra can be converted to a velocity distribution spectra
considering the proper distance covered by the jet. The velocity
distribution of a collimated supersonic jet can be expressed using the
equation:

\begin{equation}
	f(v)\ =\ a\ (v\ -\ c)^{2}\ \exp{(-b((v-u)-c)^{2}))} 
	\label{vel-dist}
\end{equation}

where \textit{a} is a normalization constant, $b\ =\ \frac{m}{2\ k\ T}$,
where \textit{m} and \textit{T} are mass and temperature of the jet, respectively; \textit{k} is the
Boltzmann constant; \textit{u} is the flow velocity and \textit{c} is just a
parameter shifting the center of the fit \cite{Cameron,haberland1985}.\\

By fitting the velocity distribution spectra of $O_{2}^{-}$ ions
for different $O_{2} : Ar$ ratios, the parallel translational
temperature of the supersonic jet (\textit{T}) can be calculated. It is also to be noted here that the peak position of the narrowed velocity distribution of the beam represents the terminal velocity of the beam. Thus by knowing the peak position of the supersonic jet, the velocity and hence the Mach number of the jet can be calculated. \\

\begin{table}
	\centering
	\caption{Table showing different fitting parameters for Equation \ref{vel-dist} for $O_{2}$ : Ar ratios 1:4, 1:9 and 1:10.}
	\label{parameters}
	\begin{tabular}{|c|c|c|c|c|c|}
		\hline
		& a & b & c & u & $R^{2}$ \\
		\hline
		1:4 & 2.024e-04 & 1.134e-04 & 5671 & -470.8 & 0.75 \\
		\hline
		1:9 & 1.352e-05 & 7.616e-05 & 6945 & -1754 & 0.52 \\
		\hline
		1:10 & 8.924e-05 & 1.531e-04 & 5782 & -587.2 & 0.67 \\
		\hline
	\end{tabular}
\end{table}

\begin{table}
	\centering
	\caption{Table showing variation in effective adiabatic index ($\gamma_{eff}$), effective mass ($M_{eff}$), velocity of sound ($C_{0}$), Mach number and jet temperature (T) with variation in argon proportion for $O_{2}$ : Ar ratios 1:4, 1:9 and 1:10.}
	\label{temp-machno}
	\begin{tabular}{|c|c|c|c|c|c|}
		\hline
		& $\gamma_{eff}$ & $M_{eff}$ (a.m.u.) & $C_{0} (\frac{m}{s})$ & M & T(K) \\
		\hline
		1:4 & 1.56 & 35.2 & 121.40 & 42.84 & 40 \\
		\hline
		1:9 & 1.58 & 37.6 & 101.84 & 50.97 & 29.69 \\
		\hline
		1:10 & 1.58 & 37.82 & 71.84 & 72.31 & 14.86 \\
		\hline
	\end{tabular}
\end{table}


The fitted curves for the velocity distribution of $O_2^-$ ions using Equation \ref{vel-dist} for $O_{2}:Ar$ ratios 1:4,
1:9 and 1:10 are compared in Figure \ref{vel-compared}. The FWHM for $O_{2}:Ar$ ratios 1:4,
1:9 and 1:10 are 108, 135 and 95, respectively. This shows that the 1:10 ratio has the lowest FWHM value implying the most narrow velocity distribution. The different fitting parameters for $ O_{2} : Ar $ ratios 1:4, 1:9 and 1:10 are given in Table \ref{parameters}. The calculated beam temperatures and Mach numbers are shown in Table \ref{temp-machno} where  $C_{0}$ denotes speed of sound in the respective $O_{2}:Ar$ medium.\\

It can be noted from Table \ref{temp-machno} that as the proportion of argon is
increased keeping the stagnant pressure $(P_{0})$ same (~5.5 bar), the
Mach number increases notably for $O_{2}:Ar$ ratios 1:4, 1:9 and
1:10.\\

The temperature of the jet also keeps reducing with the increase in
Ar concentration. In the light of this observation, we predict
that amongst the $O_{2}:Ar$ ratios 1:4, 1:9 and 1:10, the beam
having $\frac{10}{11}$ th portion of argon forms the most effective
clusters of oxygen molecules with the highest Mach number and the
lowest temperature. 

\section{Ion-yield curves}

\begin{figure}[h]
	\centering	
	\includegraphics[width=1.1\linewidth]{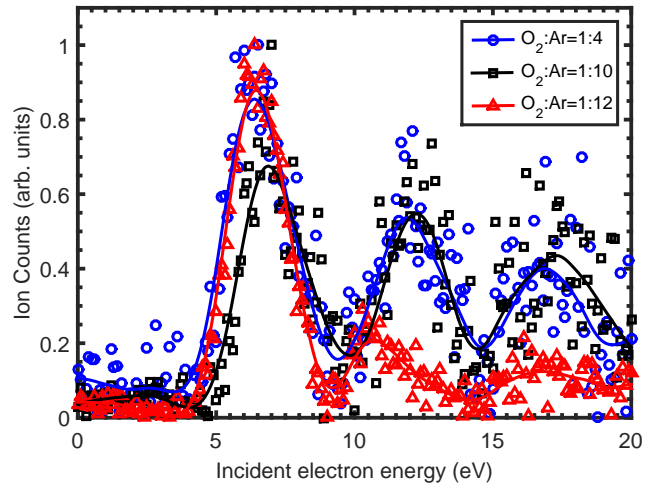}
	\caption{\label{IonO-}  Ion-yield curves of O$^-$ ions for O$_2$:Ar ratios 1:4, 1:10 and 1:12. Symbols represent experimental data and lines their correspoding fits.}
\end{figure}

\begin{figure}[h]
	\centering	
	\includegraphics[width=1.1\linewidth]{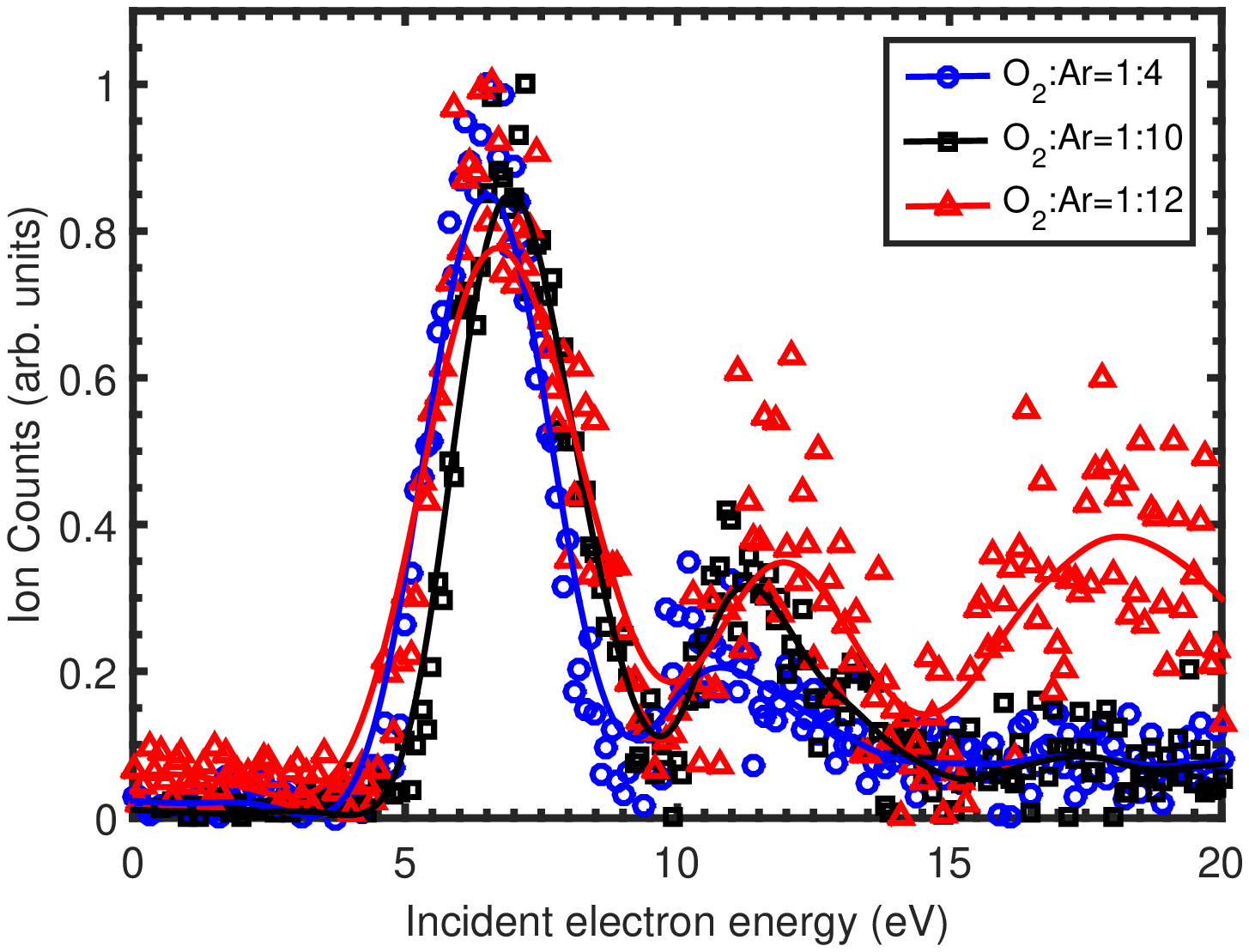}
	\caption{\label{IonO2-}  Ion-yield curves of O$_2^-$ ions for O$_2$:Ar ratios 1:4, 1:10 and 1:12. Symbols represent experimental data and lines their correspoding fits.}
\end{figure}

\begin{figure}[h]
	\centering	
	\includegraphics[width=1.1\linewidth]{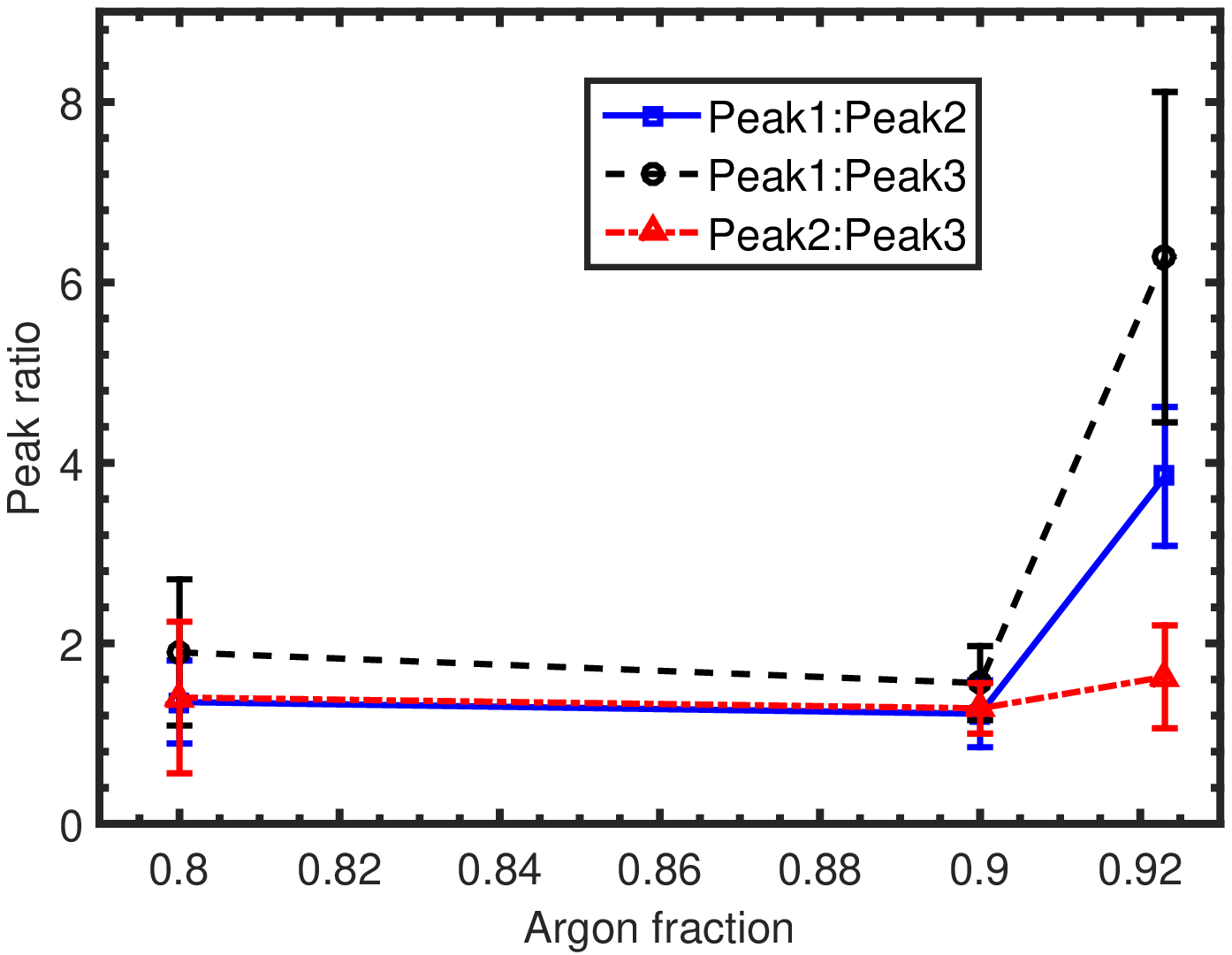}
	\caption{\label{PR-O} Ratio of different peaks with variation in Ar proportion for O$^-$ ions. The vertical lines represent error bars for each peak ratio. }
\end{figure}

\begin{figure}[h]
	\centering	
	\includegraphics[width=1.1\linewidth]{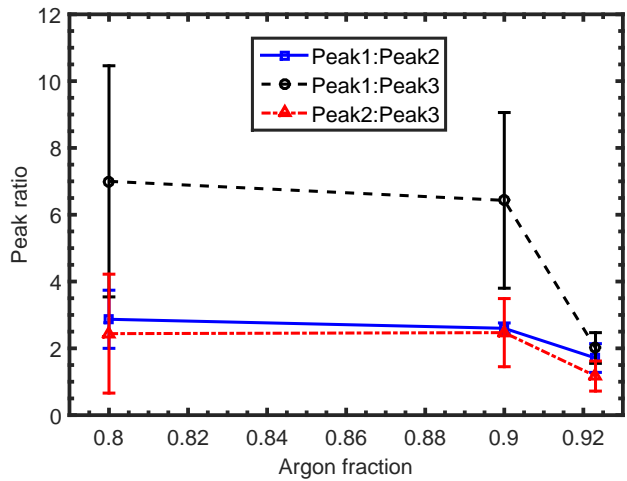}
	\caption{\label{PR-O2}  Ratio of different peaks with variation in Ar proportion for O$_2^-$ ions. The vertical lines represent error bars for each peak ratio.}
\end{figure}

The ion yield curves for $O^{-}$ and $O_{2}^{-}$ ions after proper background correction are shown in Figures \ref{IonO-} and \ref{IonO2-}, respectively for $O_{2}:Ar$
ratio 1:4, 1:10 and 1:12. The stagnant reservoir pressure was kept fixed at 5 bar, 5.5 bar and 6 bar for $O_{2}:Ar$ ratios 1:4, 1:10 and 1:12, respectively. Ion yield curve for pure
$O_{2}$ supersonic jet without any carrier shows a single resonance
peak $\sim$7 eV for both $O^{-}$ and $O_{2}^{-}$ ions. An additional peak
near 0 eV is also known to be observed by previous groups which
explains the large thermal-energy electron attachment cross-section
and low-temperature and high-pressure conditions \cite{mark1985,matejcik1996}.
At stagnation pressure 5, 5.5 and 6 bar for different $O_{2} : Ar$
ratios: 1:4, 1:10 and 1:12 , we observed two distinct new features
near $\sim$11 and $\sim$16 eV incident electron energies for both $O^{-}$ and $O_{2}^{-}$ ions as shown in the ion yield curves. The evolution of these two new features has been reported only by Illenberger previously who had observed two new peaks at $\sim$8.3 and $\sim$14.5 eV for $O^{-}$ ions at 3.5 bar stagnation pressure \cite{illenberger1992}. Illenberger also
reported that the appearance of the two new features was more
pronounced for $O_{2}^{-}$ ions at even lower stagnation pressures
(1.5 bar). While we report here some completely new observations, we
observe that the new features near $\sim$11 and $\sim$16 eV incident electron energies are comparatively much more pronounced in $O^{-}$ ion than $O_{2}^{-}$ ion for $O_{2}$ : Ar ratios 1:4 and 1:10. However, as the proportion of Argon is
increased to a even higher value of $O_{2}$:Ar = 1:12, it could be observed that the peaks near $\sim$11 and $\sim$16 eV incident electron energies become more pronounced for $O_{2}^{-}$ ion as compared to $O^{-}$ ion. \\

The appearance for these additional new features was explained by Illenberger in reference to the electron stimulated
desorption (ESD) experiments performed by Sanche \textit{et al.} \cite{illenberger1992,sanche1984}. The ground state of oxygen is a $^{3}\Sigma^{-
}_{g}$state . Electron attachment to the neutral molecule at 6.5 eV
incident electron energy occurs due to a transition of the molecule
to a $^{2}\Pi_{u}$ state. According to the $\sigma^{-}$ selection
rule, the single electron wave function should be $\sigma^{+}$ in a
single electron molecule frame of reference. Thus, transitions from
$\sigma^{-}$ to $\sigma^{+}$ states and vice-versa are not allowed as
per the $\sigma^{-}$ selection rule. But the two new resonance peaks
occurring at $\sim$11 and $\sim$16 eV incident electron energies support the
fact that transitions from $^{3}\Sigma^{-}_{g}$ to $^{2}\Sigma^{+}_{g} $ and $^{3}\Sigma^{-}_{g} $ to $^{2}\Sigma^{+}_{u}$
are giving rise to the peaks near $\sim$11 and $\sim$16 eV, respectively \cite{illenberger1992}. This points to the fact that in the condensed phase, the
$\sigma^{-}$ selection rule is violated. Evidence for the violation of
the $\sigma^{-}$ selection rule at the condensed phase of materials
have also been given previously for ESD experiments \cite{Azria-selectionrule,sanche1984}. \\

In light of these findings of the symmetry-forbidden transitions,
another explanation can be provided. The incoming electrons are
initially inelastically scattered in the cluster and slowed down.
These slowed down electrons in the vicinity of the cluster is
resonantly captured by the molecule giving rise to the negative ions.
Thus the secondary reactions may give rise to the symmetry forbidden
states in the ion-yield curves \cite{illenberger1992,schulz1962}.\\

 It can also be observed that these symmetry forbidden states are a
function of the stagnant pressure and also a function of the
proportion of the carrier gas Ar. The ratios of the second and
third resonance peaks with respect to the first peak are plotted in
Figures \ref{PR-O} and \ref{PR-O2} for $ O^{-}$ and $O^{-}_{2}$ ions, respectively.The ratios are clearly seen to follow a particular trend with the
change in the concentration of the carrier gas. The $O_{2} : Ar$
ration 1:10 gives the optimum peak ratio. Shifting the argon
proportion below or above this proportion results in considerable
decrease in the number of counts of the two new resonance peaks at
$\sim$11 and $\sim$16 eV incident electron energies. This points to the fact
that the concentration of the carrier gas has a very important role to
play in cooling down of the supersonic jet and hence the effective
formation of clusters.\\

When a beam of polyatomic molecules, seeded in a monoatomic inert
carrier gas, undergoes supersonic expansion, the translational
temperature of the carrier gas falls to an extreme low value. The
polyatomic gas molecules undergo two-body collisions with this low-
temperature bath created by the carrier gas resulting in translational
and rotational cooling of the polyatomic gas jet. The decrease in the
proportion of the seeder gas below a certain amount definitely affects
the cooling of the jet making the formation of clusters less
effective. While, if the proportion of the carrier gas is too high,
complex formation of the mono-atomic inert gas again sets a limit on
the cooling of the supersonic jet. At some point, the molecules start
colliding with the atoms and begin to polymerize giving off inter-atomic
binding energy. This again reheats the jet making the cooling less
effective \cite{smalley1977}. 

\section{Conclusion}
The present results indicate the presence of symmetry forbidden
electronic transitions in electron attachment processes to isolated
oxygen molecules in supersonic oxygen jet and also $(O_{2})_{n}$
homologous clusters. Thus DEA to $ O_{2}$ supersonic beam containing
neutral clusters can provide a method by which resonances not observed
in the gas phase can be detected. The cross-section of such forbidden
electronic transitions can also be controlled by changing the
concentration o the carrier gas. However, a time-dependent nuclear
dynamics would be helpful to understand the mechanism responsible for
such transitions.

\section{Acknowledgments}
We gratefully acknowledge financial supports from ``Science and Engineering Research Board (SERB)'' for supporting this research under the Project ``EMR/2014/000457''
\bibliography{reference_supersonic.bib}
\end{document}